\documentclass[aps,prd,reprint,superscriptaddress]{revtex4-1}
\pdfoutput=1
\usepackage{graphicx}
\usepackage{amsmath}
\usepackage{amssymb}
\newcommand{\sun}{\ensuremath{\odot}}

\begin{document}


\title{The imprint of the relative velocity between baryons and dark matter on the 21-cm signal from reionization}


\author{Jonathan M. Bittner}
\email[]{jbittner@physics.harvard.edu}
\affiliation{Jefferson Laboratory of Physics, Harvard University, 17 Oxford St, Cambridge, MA 02138}
\affiliation{Institute for Theory \& Computation, Harvard-Smithsonian CfA, 60 Garden Street, Cambridge, MA 02138}

\author{Abraham Loeb}
\email[]{aloeb@cfa.harvard.edu}
\affiliation{Institute for Theory \& Computation, Harvard-Smithsonian CfA, 60 Garden Street, Cambridge, MA 02138}

\date{\today}

\begin{abstract}

The post-recombination streaming of baryons through dark matter keeps baryons out of low mass ($<10^6 M_{\sun}$) halos coherently on scales of a few comoving Mpc. It has been argued that this will have a large impact on the 21-cm signal before and after reionization, as it raises the minimum mass required to form ionizing sources. Using a semi-numerical code, we show that the impact of the baryon streaming effect on the 21-cm signal during reionization (redshifts $z \approx$ 7-20) depends strongly on the cooling scenario assumed for star formation, and the corresponding virial temperature or mass at which stars form. For the canonical case of atomic hydrogen cooling at $10^4$ K, the minimum mass for star formation is well above the mass of halos that are affected by the baryon streaming and there are no major changes to existing predictions.  For the case of molecular hydrogen cooling, we find that reionization is delayed by a $\Delta z \approx 2$ and that more relative power is found in large modes at a given ionization fraction. However, the delay in reionization is degenerate with astrophysical assumptions, such as the production rate of UV photons by stars. 
\end{abstract}


\maketitle


\section{Introduction}
The relative velocity of baryons through cold dark matter ($v_{bc}$) after recombination is both supersonic and coherent on scales of a few comoving Mpc \cite{2010PhRvD..82h3520T,2010JCAP...11..007D,2010arXiv1012.2574T}. This has potentially interesting implications for star formation \cite{2011ApJ...730L...1S,2011arXiv1105.3732Y, 2011arXiv1101.5493G,2011MNRAS.412L..40M}.

In particular, some baryons are able to stream out and escape the potential wells of mini-halos, reducing the available supply of gas to form stars \cite{2011arXiv1108.5176N}.  Minihalos below $10^5 M_{\sun}$ will have their baryon fractions strongly suppressed, and since the effect is velocity dependent (and therefore spatially dependent), there will be rare regions of higher streaming velocity where strong suppression can occur in halos as large as $10^6 M_{\sun}$ \cite{2011arXiv1108.5176N}. Overall, the escape of baryons from dark matter potential wells can be quantified simply by assuming a characteristic ``filtering'' mass, above which baryons can be captured in dark matter halos \cite{2010arXiv1012.2574T}.

Since the $v_{bc}$ effect suppresses baryon capture by halos, one way to study it is through the 21-cm signal from hydrogen during the Epoch of Reionization (EoR).  If stars reionized the universe, their formation history will shape the history of the ionization fraction and thus the 21-cm signal. As many observing teams are presently making rapid progress towards observing or constraining the 21-cm signal at the end of reionization \cite{2010Natur.468..796B, 2011AAS...21813207W}, it is natural to ask what the imprint of the $v_{bc}$ effect will be on this measurement.

We find in this work that the result depends strongly on the assumed minimum mass for star formation.  If the minimum mass needed to cool and produce stars is higher than the suppression scale, there should be (almost) no effect on reionization. However, if the minimum mass needed to cool is below or of order the suppression scale, the $v_{bc}$ effect will raise the minimum mass to the suppression scale in a position-dependent way, and will couple the $v_{bc}$ power spectrum to the 21-cm power spectrum. The traditional filtering mass for star formation is set by the linear theory evolution history, and thus it is dominated at low redshift ($z<150$) by adiabatic cooling \cite{2007MNRAS.377..667N}. The virial temperature of a halo depends on its mass \cite{LoebText}, $\mu$, and redshift z:

\begin{equation}
T_{vir} = 1.04 \times 10^4 \left(\frac{\mu}{0.6}\right) \left(\frac{M}{10^8 M_{\sun}}\right)^{2/3} \left(\frac{1+z}{10}\right) {\rm K},
\end{equation} where $\mu$ is the mean molecular weight. Since virial temperature increases monotonically with halo mass, it is possible to convert a characteristic cooling temperature into a redshift dependent minimum mass that can be compared to the filtering mass: \cite{2011MNRAS.411..955M, 2011ApJ...730L...1S, LoebText, 2011arXiv1110.2111F}.

\begin{equation}
M_{min} = 10^8 M_{\sun} \left[ \left( \frac{T_{min}}{10^4 {\rm K}} \right) \left(\frac{0.6}{\mu}\right) \left(\frac{10}{1+z}\right) \right]^{3/2}.
\end{equation}

An important unanswered question is what form of gas cooling is responsible for producing the stellar photons which reionized the universe. Molecular hydrogen cooling at $300$ K should occur in primordial gas, leading to the formation of metal-free Population III stars, but the Lyman-Werner (LW) UV background from those stars will also rapidly photo dissociate molecular hydrogen throughout the universe \cite{1997ApJ...476..458H, 2000ApJ...534...11H}. In the absense of $\rm H_2$, atomic hydrogen cooling at $10^4$ K would be most relevant, and would set the minimum halo mass for star formation at $M_{min} = 10^8 M_{\sun} [(1+z)/10)]^{-3/2}$ \cite{2011arXiv1105.5648H}.

Radiative feedback makes it difficult to know whether atomic or molecular cooling was dominant for the first stars and galaxies. The dynamic range of scales involved is difficult to simulate, and there are many aspects of star formation in the early universe that remain unclear \cite{2010ApJ...712..101W}. The relative importance of the cooling types may be spatially dependent \cite{2011arXiv1105.5648H}, leading in principle to a complex interplay of spatially varying minimum masses from both the $v_{bc}$ effect and the spatially dependent LW background. Reference \cite{2011arXiv1105.5648H}, for example, finds that the first halos which form through molecular cooling will destroy their own reserviors of $\rm H_2$. This makes atomic cooling more relevant to the minimum mass during the reionization epoch, so that the spatially-varying effect of molecular hydrogen can be ignored. Some recent studies suggest that molecular hydrogen will dissociate quickly and that ``self-shielding'' in some regimes is even less important than previously thought \cite{2011arXiv1106.3523W}.

These recent results support the notion that the atomic cooling scenario is more likely to be dominant during the reionization epoch than molecular cooling, which in turn will decrease the impact of the $v_{bc}$ effect on the 21-cm signal. However, a considerable body of previous work has emphasized the importance of the $v_{bc}$ effect to the 21-cm signal using calculations based on molecular cooling \cite{2010JCAP...11..007D, 2010PhRvD..82h3520T}. While it is valuable to consider the impact of the $v_{bc}$ in its most extreme incarnation, we consider the likely impact of the $v_{bc}$ effect as a function of our uncertainty about the cooling scenario, so that the cases may be discriminated from each other.

\begin{figure}[h!]
\includegraphics[width=3.25in]{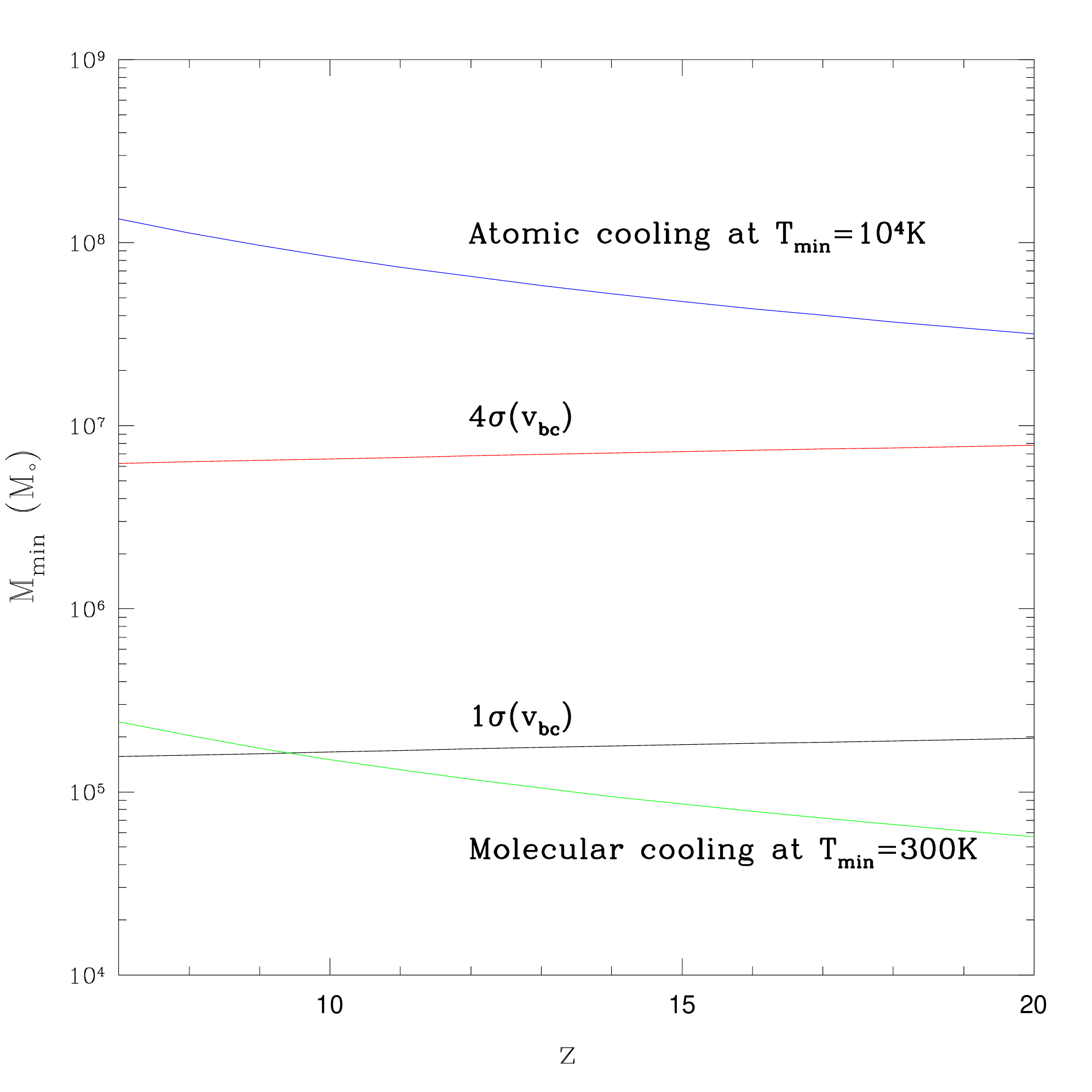}
\caption{\label{Fig 1} The evolution of the minimum mass created by the $v_{bc}$ effect, at one and four times the rms velocity of the $v_{bc}$ ($\sigma$) (from Ref. \cite{2010arXiv1012.2574T}), compared to the minimum mass for star formation associated with molecular and atomic hydrogen cooling.}
\end{figure}

\section{Method of calculation}

The $v_{bc}$ streaming is coherent over a few comoving Mpc (cMpc) scales while star formation occurs at scales many orders of magnitude smaller. We study the imprint of the effect with a semi-numerical code {\it 21cm FAST} \cite{2011MNRAS.411..955M} with a 400 cMpc box with 4 cMpc resolution. Semi-numerical codes like  {\it 21cm FAST} realize random cosmological initial conditions, and use linear pertubation theory and analytic prescriptions for the relevant physics to evolve the volume over time. This volume can be made large enough to represent a fair sample of the universe.

We modified {\it 21cm FAST} in order to incorporate the $v_{bc}$ effect by self-consistently generating a coherent $v_{bc}$ velocity in each 4 cMpc voxel. We then modify the minimum mass locally in that voxel according to the prescription set by the filtering masses in Ref \cite{2010arXiv1012.2574T}. Our study is conducted in the regime where $T_s \gg T_{CMB}$, which is valid for $z \le 20$ \cite{2011arXiv1109.6012P}. This greatly simplifies our analysis because we do not have to consider the inhomogenous heating of the IGM by the first soures and the effect of the $v_{bc}$ on it. We assume the default cosmological parameters for {\it 21cm FAST} with no halo finding.

\begin{figure}[h!]
\includegraphics[width=3.25in]{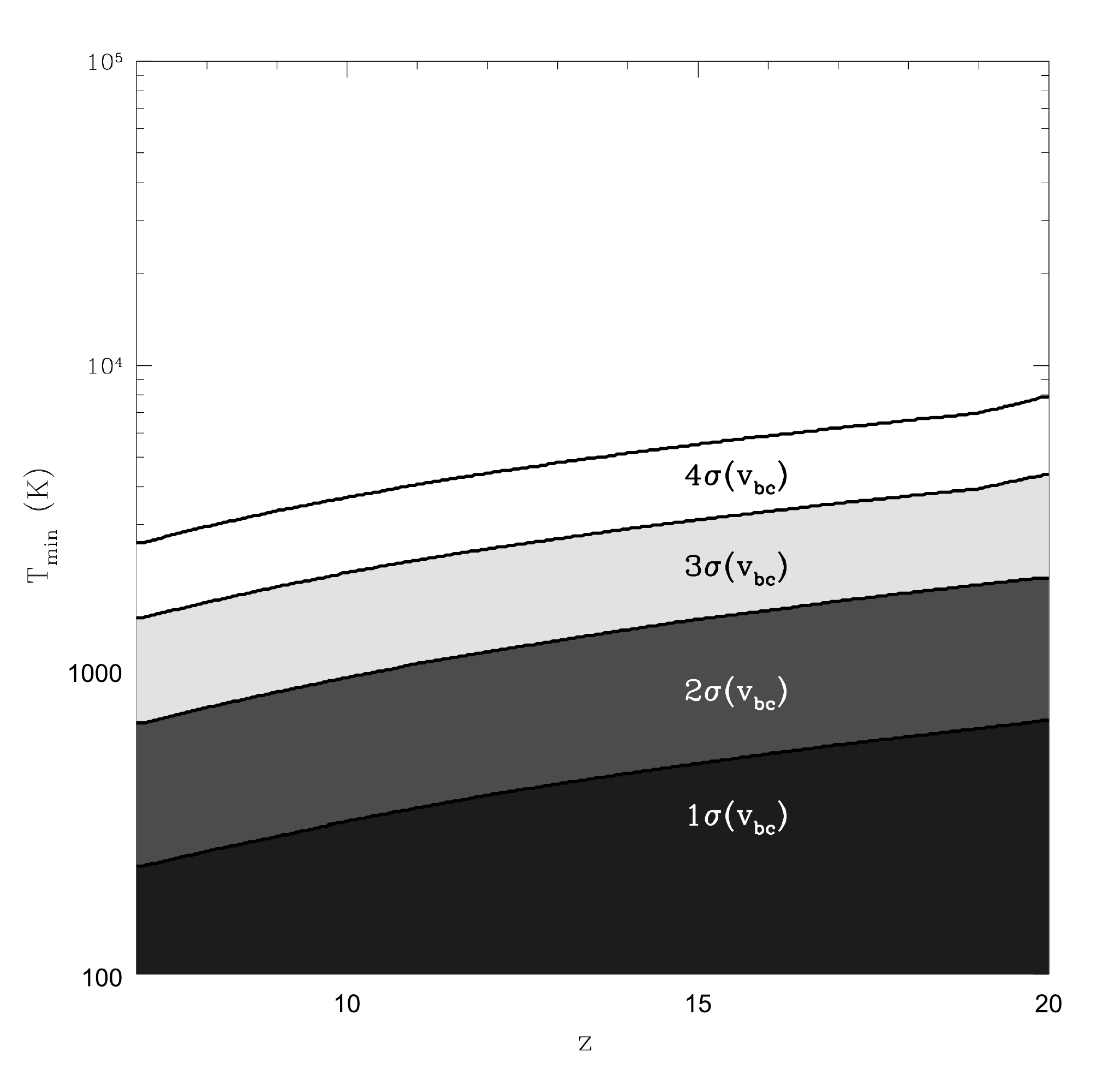}
\caption{\label{Fig 2}The magnitude of the $v_{bc}$ effect needed to modify the minimum mass of ionizing halos, giving a particular cooling temperature, as a function of redshift. The shaded regions correspond to 1$\sigma$, 2$\sigma$, 3$\sigma$, and 4$\sigma$ fluctuations of the $v_{bc}$ effect, from bottom to top.}
\end{figure}

We can then run realizations of the universe with the $v_{bc}$ effect turned both on and off, for four different fiduciual cooling scenarios (and therefore minimum masses). The four scenarios involve minimum virial temperatures, $T_{min}$, of: 200K, 300K, $10^4$K, and $10^5$K.

We obtain frequency and redshift dependent 21-cm differential brightness power spectrum, images, and mean ionization fractions as a function of redshift from {\it 21-cm FAST}. Our results are described below.

\section{Results}

We wish to find the redshifts at which the new filtering mass implied by the $v_{bc}$ effect raises the minimum mass implied by the cooling scenario under consideration. In Figure 1, we show the redshift at which the minimum masses implied by molecular and atomic cooling become modified by a 1-sigma and 4-sigma fluctuation of the $v_{bc}$. We also show in Figure 2 which regions of minimum-mass and cooling temperature parameter space are affected by the $v_{bc}$ effect at different multiples of the rms velocity.

\begin{figure}[t!]
\includegraphics[width=3.25in]{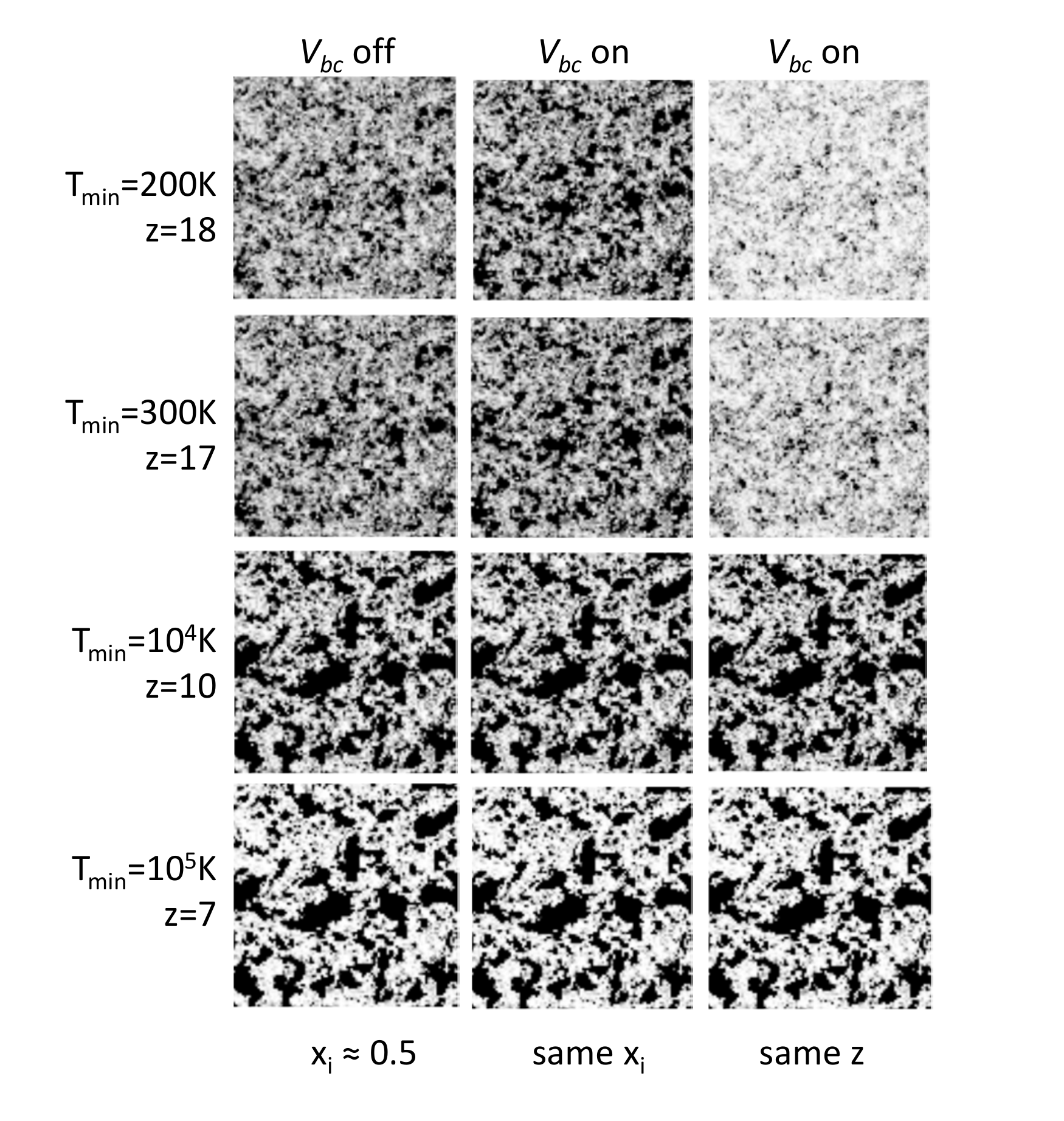}
\caption{\label{Fig 3}Images of ionization fraction, 400Mpc on a side, with and without the $v_{bc}$ effect. White represents neutral gas and black represents ionized gas. The first column has no $v_{bc}$ effect and is shown at a fixed $x_i = 0.5$, with each row using a different cooling temperature and redshift. The second column uses the same cooling temperatures as the first column, but includes the $v_{bc}$ effect and shows the equivalent panel at the same $x_i=0.5$ (which for the top two rows is at a different redshift due to the delay in reionization). The third column also includes the $v_{bc}$ effect at the same cooling temperatures, but shows the image at the same redshift as the left column (which in the first two rows, is at a different ionization fraction). At higher cooling temperatures / minimum masses, there is essentially no difference in the three panels.}
\end{figure}

To study the effect of the inhomogeneous $v_{bc}$ effect, we realize eight different cosmological histories using a modified version of {\it 21-cm FAST}, and compare the results as a function of minimum mass and by turning the $v_{bc}$ effect on and off. Figure 3 shows image slices of the ionization fields for the both cases, and compares how the $v_{bc}$ effect modifies the image both at an equivalent redshift and at the same ionization fraction $x_i = 0.5$. As can be seen by comparing the left column images to the images in the right two columns, the $v_{bc}$ effect only imprints onto the ionization field  if molecular cooling is important, in which case it increases the typical size of the neutral voids and delays reionization. Figure 4 shows the difference in power spectra due to the $v_{bc}$ effect, which clearly shows the transfer of power to longer wavelength modes. Figure 5 shows the global delay in reionization due to the overall higher effective minimum mass, in agreement with the results of Ref \cite{2010JCAP...11..007D}.

\begin{figure}[h!]
\includegraphics[width=3.0in]{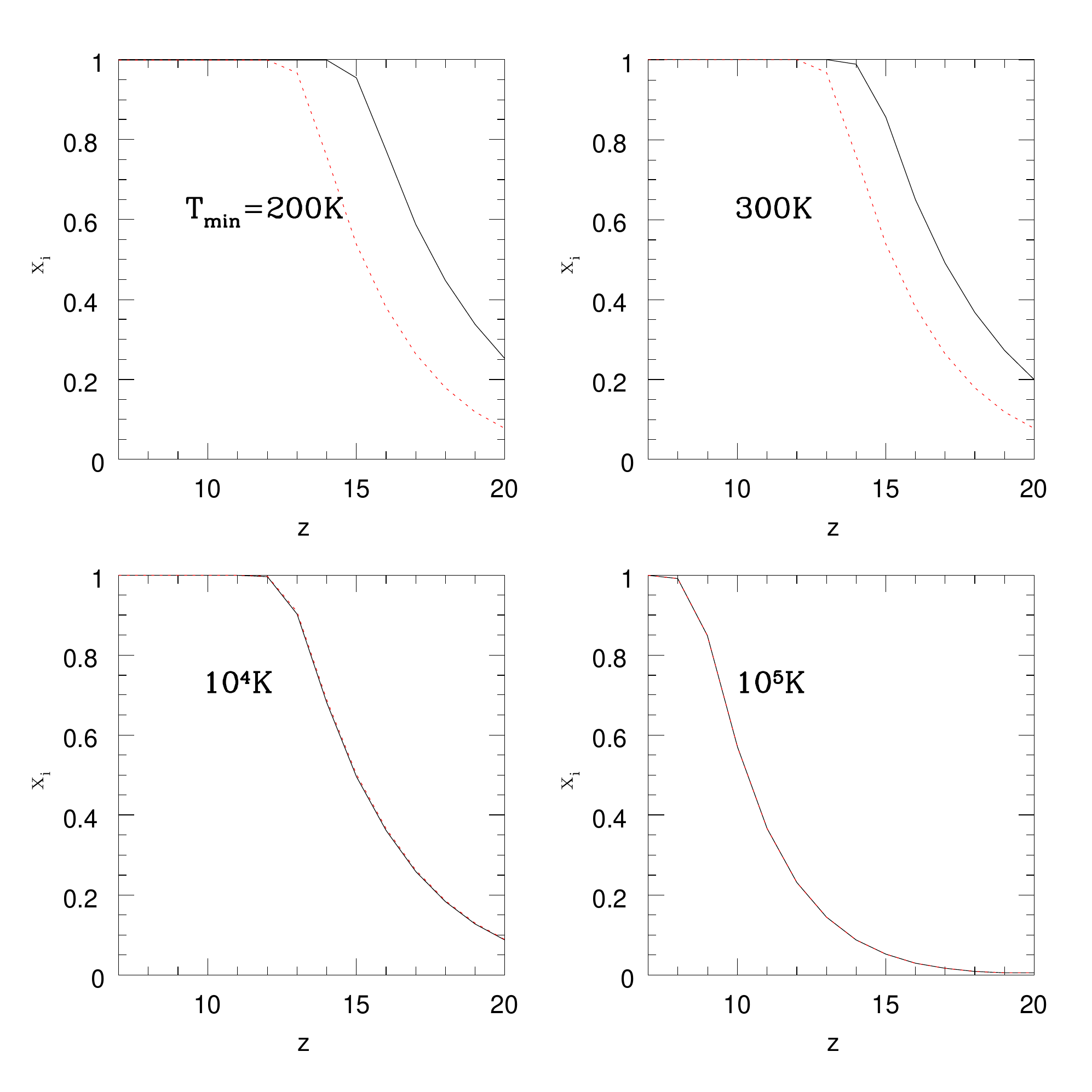}
\caption{\label{Fig 4}Ionization fraction history as a function of redshift, with and without the $v_{bc}$ effect, with a minimum halo mass for star formation associated with cooling at 200K, 300K, $10^4$K, and $10^5$K (top left, top right, bottom left, and bottom right respectively). The dashed line includes the $v_{bc}$ effect while the solid line does not include the $v_{bc}$ effect. Other parameters are {\it21cm FAST} defaults ($f_{*} = 0.1$, $\xi^{-1} = 31.5$).}
\end{figure}

\begin{figure}[h!]
\includegraphics[width=3.0in]{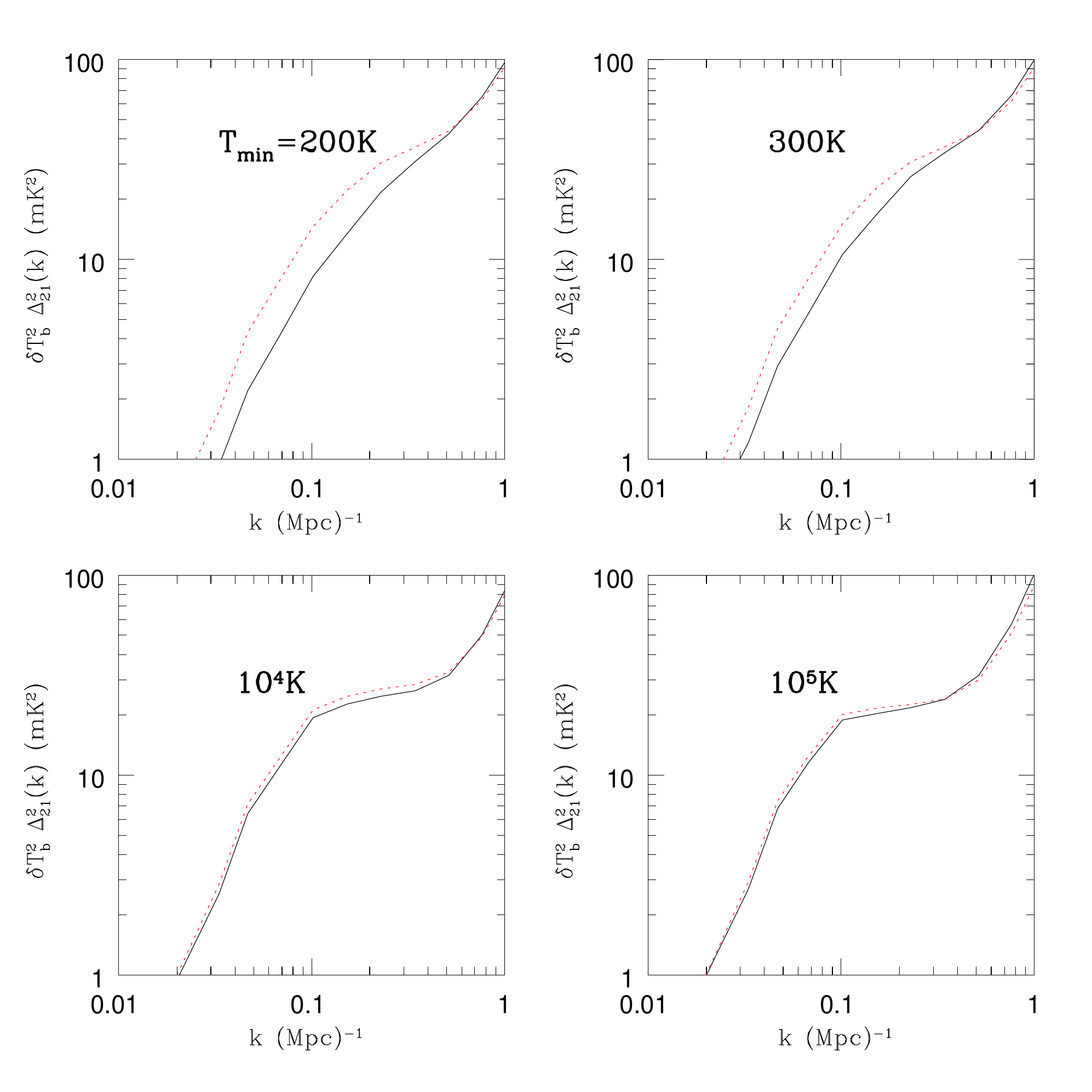}
\caption{\label{Fig 5}The 21-cm differential brightness temperature $\delta T_b$ power spectrum, with and without the $v_{bc}$ effect, at fixed $x_{i}$=0.5, with a minimum halo mass for star formation associated with cooling at 200K, 300K, $10^4$K, and $10^5$K (top left, top right, bottom left, and bottom right respectively).The dashed line includes the $v_{bc}$ effect while the solid line does not include the $v_{bc}$ effect.}
\end{figure}

\section{Conclusion}

Given the likely scenario of atomic hydrogen cooling, the $v_{bc}$ effect at $z \le 20$ will probably not be detectable during the Epoch of Reionization. Molecular cooling may be dominant at $z > 20$, but in our fiducial model, sources which produce most of the UV photons causing reionization are formed at $z \le 20$. It is very challenging to detect the corresponding 21-cm signal at $z > 20$, because the galactic synchrotron foreground has a brightness temperature $T_b$ that rises steeply with observed frequencies $\nu_{obs}$, $T_b \propto \nu_{obs}^{-2.6}$. 

The regime $z \gtrsim 20$ also coincides with the time at which the first sources started heating the spin temperature above the CMB temperature, making the 21-cm signal both more interesting and harder to compute. However, the assumptions about cooling (and thus minimum mass) strongly modulate the $v_{bc}$ effect and future work should study how $v_{bc}$ predictions change in the high redshift regime as a function of minimum mass prescriptions.

If molecular cooling does turn out to be dominant for sources that produce most of the UV photons, the $v_{bc}$ effect will delay reionization. However, this delay will be degenerate with uncertainties in the star-forming efficiency $f_{*}$  the escape fraction of ionizing photons from galaxies $f_{esc}$, and other astrophysical parameters such as the mass function of stars.  We find that the shape of the power spectrum predicted by {\it 21-cm FAST} changes in a characteristic way from the no-$v_{bc}$ case (at a fixed ionization fraction). Our ability to measure or constrain the $v_{bc}$ effect in this case will depend on whether we can use simulations and semi-numerical models to reliably and accurately predict the 21-cm signal.

\section*{Acknowledgments}

We thank Smadar Naoz for helpful comments. This work was supported in part by NSF grants AST-0907890, and NASA
grants NNX08AL43G and NNA09DB30A.

\def\aj{AJ}                   
\def\araa{ARA\&A}             
\def\apj{ApJ}                 
\def\apjl{ApJ}                
\def\apjs{ApJS}               
\def\ao{Appl.~Opt.}           
\def\apss{Ap\&SS}             
\def\aap{A\&A}                
\def\aapr{A\&A~Rev.}          
\def\aaps{A\&AS}              
\def\azh{AZh}                 
\def\baas{BAAS}               
\def\jrasc{JRASC}             
\def\memras{MmRAS}            
\def\mnras{MNRAS}             
\def\pra{Phys.~Rev.~A}        
\def\prb{Phys.~Rev.~B}        
\def\prc{Phys.~Rev.~C}        
\def\prd{Phys.~Rev.~D}        
\def\pre{Phys.~Rev.~E}        
\def\prl{Phys.~Rev.~Lett.}    
\def\pasp{PASP}               
\def\pasj{PASJ}               
\def\qjras{QJRAS}             
\def\skytel{S\&T}             
\def\solphys{Sol.~Phys.}      
\def\sovast{Soviet~Ast.}      
\def\ssr{Space~Sci.~Rev.}     
\def\zap{ZAp}                 
\def\nat{Nature}              
\def\iaucirc{IAU~Circ.}       
\def\aplett{Astrophys.~Lett.} 
\def\apspr{Astrophys.~Space~Phys.~Res.}
\def\bain{Bull.~Astron.~Inst.~Netherlands} 
\def\fcp{Fund.~Cosmic~Phys.}  
\def\gca{Geochim.~Cosmochim.~Acta}   
\def\grl{Geophys.~Res.~Lett.} 
\def\jcp{J.~Chem.~Phys.}      
\def\jgr{J.~Geophys.~Res.}    
\def\jqsrt{J.~Quant.~Spec.~Radiat.~Transf.}
\def\memsai{Mem.~Soc.~Astron.~Italiana}
\def\nphysa{Nucl.~Phys.~A}   
\def\physrep{Phys.~Rep.}   
\def\physscr{Phys.~Scr}   
\def\planss{Planet.~Space~Sci.}   
\def\procspie{Proc.~SPIE}   

\let\astap=\aap
\let\apjlett=\apjl
\let\apjsupp=\apjs
\let\applopt=\ao

\let\astap=\aap
\let\apjlett=\apjl
\let\apjsupp=\apjs
\let\applopt=\ao

\bibliography{vbcbib6}

\end{document}